\documentclass[aps,twocolumn,showpacs,floatfix,groupedaddress,amsmath,amssymb,pra]{revtex4}
\usepackage{graphicx}
\usepackage{mathrsfs}
\usepackage{array}


\newcommand{\ket} [1] {| #1 \rangle}
\newcommand{\bra} [1] {\langle #1 |}
\newcommand{\braket}[2]{\langle #1 | #2 \rangle}
\newcommand{\ketbra}[2]{\ket{#1}\bra{#2}}
\newcommand{\proj}[1]{\mbox{$|#1\rangle \!\langle #1 |$}}
\makeatletter

\newcommand{\eref}[1]{Eq.~\ref{#1}}

\newcommand{\fref}[1]{Fig.~\ref{#1}}

\newcommand{\markend}{~\rule{0.4em}{1.8ex}} 
\newcommand{\tr}{\mbox{tr}}
\newcommand{\nsym}{\tiny \mbox{sym}}
\newcommand{\nmin}{\tiny \mbox{min}}

\usepackage[colorlinks]{hyperref}

\usepackage{array}

\begin{document}
\title
{Global symmetries in tensor network states: \\ symmetric tensors versus minimal bond dimension}
\author{Sukhwinder Singh}
\affiliation{Center for Engineered Quantum Systems; Dept. of Physics \& Astronomy, \\Macquarie University, 2109 NSW, Australia}
\author{Guifre Vidal}
\affiliation{Perimeter Institute for Theoretical Physics, Waterloo, Ontario, N2L 2Y5, Canada}

\begin{abstract}
Tensor networks offer a variational formalism to efficiently represent wave-functions of extended quantum many-body systems on a lattice. In a tensor network $\mathcal{N}$, the dimension $\chi$ of the bond indices that connect its tensors controls the number of variational parameters and associated computational costs. In the absence of any symmetry, the \textit{minimal} bond dimension $\chi^{\nmin}$ required to represent a given many-body wave-function $\ket{\Psi}$ leads to the most compact, computationally efficient tensor network description of $\ket{\Psi}$. In the presence of a global, on-site symmetry, one can use a tensor network $\mathcal{N}_{\nsym}$ made of \textit{symmetric} tensors. Symmetric tensors allow to exactly preserve the symmetry and to target specific quantum numbers, while their sparse structure leads to a compact description and lowers computational costs.
In this paper we explore the trade-off between using a tensor network $\mathcal{N}$ with minimal bond dimension $\chi^{\nmin}$ and a tensor network $\mathcal{N}_{\nsym}$ made of symmetric tensors, where the minimal bond dimension $\chi^{\nmin}_{\nsym}$ might be larger than $\chi^{\nmin}$. We present two technical results. First, we show that in a tree tensor network, which is the most general tensor network \textit{without loops}, the minimal bond dimension can always be achieved with symmetric tensors, so that $\chi^{\nmin}_{\nsym} = \chi^{\nmin}$. Second, we provide explicit examples of tensor networks \textit{with loops} where replacing tensors with symmetric ones necessarily increases the bond dimension, so that $\chi_{\nsym}^{\nmin}>\chi^{\nmin}$. We further argue, however, that in some situations there are important conceptual reasons to prefer a tensor network representation with symmetric tensors (and possibly larger bond dimension) over one with minimal bond dimension.
\end{abstract}


\maketitle
\section{Introduction}

Tensor networks allow for an efficient description of certain quantum many-body wave-functions, including the ground states of a large variety of local Hamiltonians on a lattice in $D$ spatial dimensions. Tensor network states were originally proposed as the basis for novel variational approaches to quantum many-body problems \cite{Fannes92, Ostlund95, Vidal03, PerezGarcia07, Shi06, Tagliacozzo09, Murg10, Vidal07b, Vidal08, Verstraete04}, and are currently used to study e.g. systems of frustrated antiferromagnets \cite{Fr1, Fr2, Fr3, Fr4, Fr5, Fr6, Fr7, Fr8} and strongly interacting fermions \cite{Fe1, Fe2, Fe3, Fe4, Fe5, Fe6, Fe7, Fe8, Fe9, Fe10, Fe11}, as well as quantum phase transitions \cite{QPT1, QPT2, QPT3, QPT4, QPT5} and topological order \cite{To1, To2, To3, To4, To5, To6, To7, To8, To9}. In addition, they have in time also become a natural language in which to formulate important theoretical questions. For instance, they have been recently used to obtain a complete classification of the possible zero-temperature, gapped phases of quantum matter in $D=1$ spatial dimensions \cite{Pollmann, Chen1, Schuch11,Chen2} and in generalizations of this result to $D>1$ spatial dimensions \cite{Chen3, Chen4, Gu12}; or adopted as a new tool to describe the holographic principle by the string theory community \cite{Maldacena, Witten, Swingle1, Swingle2, Branching, Metric, Ho1, Ho2, Ho3, Ho4, Ho5}.

\subsection{Tensor networks}

The wave-function $\ket{\Psi} \in \mathbb{V}^{\otimes L}$ of a many-body system on a lattice $\mathcal{L}$ made of $L$ sites, each described by a $d$-dimensional vector space $\mathbb{V}$, can always be expressed as a linear combination of $d^{L}$ product basis states $\ket{i_1 i_2\ldots i_L} \equiv \ket{i_1} \otimes \ket{i_2} \otimes \cdots \otimes \ket{i_L}$,
\begin{equation}
\ket{\Psi} = \sum_{i_1=1}^d \sum_{i_2=1}^d \cdots \sum_{i_L=1}^d \hat{\Psi}_{i_1i_2\ldots i_{L}} \ket{i_1i_2\ldots i_L}.\label{eq:genstate}
\end{equation}
Here $\{\ket{i_k}\}_{i_k=1}^{d}$ is an orthonormal basis of the vector space $\mathbb{V}$ of site $k$, and $\hat{\Psi}_{i_1i_2\ldots i_{L}}$ are complex coefficients.

\begin{figure}[t]
  \includegraphics[width=6.5cm]{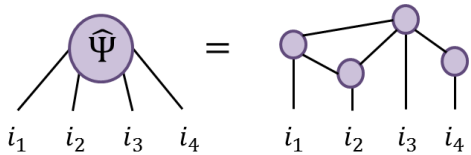}
\caption{
Graphical representation of a tensor network $\mathcal{N}$ for a many-body wave-function $\ket{\Psi}$ of a lattice $\mathcal{L}$ made of $L=4$ sites. Each circle represents a tensor and the emerging lines depict the indices of the tensor. Each \textit{bond} index of the tensor network connects two tensors. There are also $L=4$ \textit{physical} indices. Bond indices are contracted (or summed over). Upon contraction of the bond indices, the tensor network produces the complex coefficients $\hat{\Psi}_{i_1i_2i_3i_4}$.   \label{fig:TN}}
\end{figure}

A tensor network representation $\mathcal{N}$ of the many-body wave-function $\ket{\Psi}$ consists of a set of tensors that are interconnected into a network, see \fref{fig:TN}. The tensor network has two types of indices: physical indices and bond indices. A \textit{physical index} (or open index) is associated to one of the $L$ sites of lattice $\mathcal{L}$ and has dimension $d$ (that is, it labels $d$ options, namely the elements $\ket{i_k}$ of a basis in $\mathbb{V}$). A \textit{bond index} connects two tensors in the network and has dimension $\chi$, the \textit{bond dimension} of the network. [For notational simplicity, in several parts of the paper we will assume $\chi$ to be uniform throughout the network, although all of our results are also valid when each bond index is allowed to have a different bond dimension.] Upon summing over (or \textit{contracting}) all bond indices, the tensor network produces a single tensor with $L$ open indices, corresponding to the $d^L$ complex coefficients $\hat{\Psi}_{i_1i_2\ldots i_{L}}$. The key point is that, by restricting the value of the bond dimension $\chi$, the tensor network represents certain many-body wave-functions $\ket{\Psi}\in \mathbb{V}^{\otimes L}$ using a number of coefficients that only grows polynomially (and often just linearly) in $L$, instead of exponentially in $L$, and thus offers an efficient description of a subset of many-body states.

Some popular classes of tensor networks, characterized primarily by the shape of the networks, have acquired special names. For instance, when the tensors are connected into a one-dimensional array, the tensor network is known as a \textit{matrix product state} \cite{Fannes92,Ostlund95,Vidal03,PerezGarcia07} (MPS), whereas when they are connected into a tree, it is known as a \textit{tree tensor network} \cite{Shi06, Tagliacozzo09, Murg10} (TTN), see \fref{fig:TTNMPS}. Notice that an MPS is a particular case of a TTN, and that the latter is characterized by having no loops. In contrast, other tensor networks have plenty of loops. This is the case of the \textit{multi-scale entanglement renormalization ansatz} \cite{Vidal07b,Vidal08} (MERA), described by a quantum circuit in $D+1$ dimensions ($D\geq 1$), and the \textit{projected entangled pair state} \cite{Verstraete04} (PEPS), where tensors are connected according to a $D$-dimensional array ($D\geq 2$), see \fref{fig:MERAPEPS}.

\subsection{Minimal bond dimension}

Given a choice of tensor network, say an MPS, and a value $\chi$ of its bond dimension, the tensor network can represent some subset $\mathcal{S}_{\chi}\subset \mathbb{V}^{\otimes L}$ of states $\ket{\Psi}$ of the lattice by changing the variational parameters stored in the tensors. By increasing the bond dimension from $\chi$ to $\chi'$, and thus specifying a larger number of variational parameters, one can represent a larger subset of states, $\mathcal{S}_{\chi} \subseteq \mathcal{S}_{\chi'}$. However, increasing the bond dimension also results in a description in terms of larger tensors. This implies larger storage costs, as well as larger computational costs to manipulate the tensor network (the cost of tensor network algorithms typically scales as $O(\chi^p)$ for some power $p$, which ranges from $p=3$ for a simple MPS to $p\approx 10-20$ for MERA and PEPS in $D\geq 2$ spatial dimensions). Consequently, given a state $\ket{\Psi} \in \mathbb{V}^{\otimes L}$, it is important to identify the minimal bond dimension $\chi^{\nmin}$ capable of representing it, since this will typically lead to the most compact description and lowest computational cost.

\subsection{Global symmetry}

Let us now consider a many-body state $\ket{\Psi\nsym}\in \mathbb{V}^{\otimes L}$ that is invariant under the action of a global, on-site symmetry $\mathcal{G}$, acting on the lattice $\mathcal{L}$. This could correspond, for instance, to particle number parity conservation, ($\mathcal{G}=$ Z$_2$), particle number conservation ($\mathcal{G} = $ U$(1)$), or to spin isotropy ($\mathcal{G} = $ SU$(2)$). In this paper we refer to a state invariant under $\mathcal{G}$ such as $\ket{\Psi_{\nsym}}$ as a $\mathcal{G}$-symmetric state.

A possible way to represent a $\mathcal{G}$-symmetric state $\ket{\Psi\nsym}$ is by using a tensor network $\mathcal{N}_{\nsym}$ made of $\mathcal{G}$-symmetric tensors -- that is, where each tensor is invariant under $\mathcal{G}$ -- after the action of $\mathcal{G}$ has been conveniently extended to all the bond indices). A $\mathcal{G}$-symmetric tensor has a well-understood inner structure that allows for a compact form of storage and for faster tensor manipulations. In addition, use of $\mathcal{G}$-symmetric tensors in $\mathcal{N}_{\nsym}$ has several other advantages, such as ensuring that the symmetry is exactly preserved even during an approximate computation, and allowing to target a specific symmetry sector (for instance, one could choose to target the \textit{odd} parity sector for $\mathcal{G}=$ Z$_2$, the sector with a given number $N=7$ of particles for $\mathcal{G} = $ U$(1)$, or sector with spin $J=1/2$ for $\mathcal{G} = $ SU$(2)$). As a result, on-site global symmetries have been systematically incorporated \cite{McCulloch08,Singh10,PerezGarcia10,Zhao10,Singh09,Singh11,Singh12, Weichselbaum} into tensor network algorithms by using symmetric tensors.
 
Notice, however, that a $\mathcal{G}$-symmetric state $\ket{\Psi_{\nsym}}$ can also be represented with a tensor network $\mathcal{N}$ made of non-symmetric tensors -- all we need is the global symmetry of $\ket{\Psi_{\nsym}}$, which does not require individual tensors to also be symmetric. And it may well be the case that the minimal bond dimension $\chi^{\nmin}$ is only achieved by a tensor network $\mathcal{N}$ with non-symmetric tensors. That is, it may be the case that forcing the global symmetry of $\ket{\Psi_{\nsym}}$ into each single tensor in $\mathcal{N}_{\nsym}$ results in a bond dimension $\chi_{\nsym}^{\nmin}$ larger than is actually needed. A main goal of the present paper is to establish when this may occur.

\subsection{Summary of results}

\begin{figure}[t]
  \includegraphics[width=6.5cm]{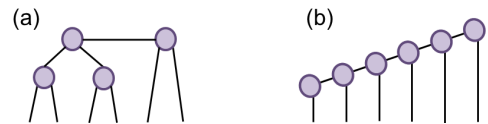}
\caption{
(a) Example of a tree tensor network (TTN) for a lattice of $L=6$ sites. Notice that the bond indices do not close any loop.
(b) A matrix product state (MPS) with open boundary conditions (OBC) is an example of TTN.
\label{fig:TTNMPS}}
\end{figure}

Given a tensor network $\mathcal{N}$ for a $\mathcal{G}$-symmetric many-body wave-function $\ket{\Psi_{\nsym}}$, we will show the following. (i) If the network contains no loops (that is, for a TTN or an MPS with open boundary conditions, see \fref{fig:TTNMPS}), then each tensor in $\mathcal{N}$ can be replaced with a symmetric tensor without increasing the bond dimension. In other words, the minimal bond dimension $\chi^{\nmin}$ for $\ket{\Psi_{\nsym}}$ can always be achieved with a tensor network $\mathcal{N}_{\nsym}$, so that $\chi^{\nmin}_{\nsym}=\chi^{\nmin}$. (ii) Instead, if the network contains one or several loops, it might be that replacing non-symmetric tensors for symmetric ones necessarily requires increasing the bond dimension, that is  $\chi^{\nmin}_{\nsym} > \chi^{\nmin}$, as we show with explicit examples for an MPS with periodic boundary conditions on a three-site lattice, and for the MERA in $D=1$ spatial dimensions.

\begin{figure}[t]
  \includegraphics[width=8.0cm]{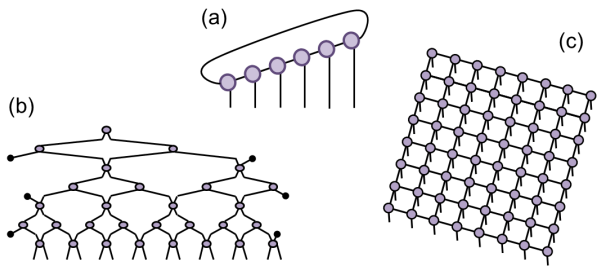}
\caption{
Examples of tensor networks with loops:
(a) matrix product state (MPS) with periodic boundary conditions (PBC) for a lattice of $L=6$ sites;
(b) multi-scale entanglement renormalization ansatz (MERA) for a lattice in $D=1$ spatial dimensions;
(c) projected entangled-pair state (PEPS) for a lattice in $D=2$ spatial dimensions.
\label{fig:MERAPEPS}}
\end{figure}

These results imply that when looking for the most compact and computationally efficient, \textit{exact} representation of $\ket{\Psi_{\nsym}}$ in terms of a TTN (or MPS with OBC), \fref{fig:TTNMPS}, we can restrict our attention to a tensor network $\mathcal{N}_{\nsym}$ made of symmetric tensors. Instead, when using tensor networks with loops, such as an MPS with PBC, a MERA or a PEPS, \fref{fig:MERAPEPS}, one still needs to consider the computational benefits of the inner structure of $\mathcal{G}$-symmetric tensors before being able to decide what type of tensors, symmetric or non-symmetric, provide the most compact and computationally efficient description.

Moreover, there are important reasons to use a tensor network $\mathcal{N}_{\nsym}$ made of $\mathcal{G}$-symmetric tensors, regardless of the above considerations, at least in two situations: in the study of symmetry-protected phases of quantum matter, and in the study of holography. Indeed, as we recently explained in Ref. \cite{SinghVidal}, the use of $\mathcal{G}$-symmetric tensors in the MERA is key to obtaining the right structure of renormalization group fixed points in the case of symmetry protected phases (see also \cite{Chen5}); and to realizing a property of the AdS/CFT correspondence, namely that a global symmetry of the boundary theory corresponds to a gauge symmetry in the bulk.

The rest of the paper is divided into four more sections. Sect. \ref{sec:invtn} reviews the use of symmetric tensors in tensor networks. Sects. \ref{sec:ttn} and \ref{sec:tnloops} present our results for tensor networks without and with loops, respectively. Finally, Sect. \ref{sec:holo} reviews the results of Ref. \cite{SinghVidal} concerning the need for $\mathcal{G}$-symmetric tensors in the MERA when applied to classifying phases and holography.

\begin{figure}[t]
  \includegraphics[width=6.5cm]{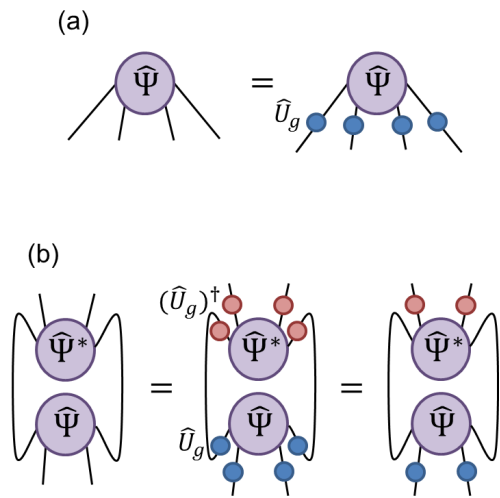}
\caption{
(a) Graphical representation of the constraint in Eq. \ref{eq:invstate} fulfilled by a $\mathcal{G}$-symmetric state $\ket{\Psi_{\nsym}}$.
(b) Graphical representation of the proof of lemma 1, or Eq. \ref{eq:invrho}. First (left) we find a representation of $\hat{\rho}^{\mathcal{A}}$ as a partial trace of $\ket{\Psi_{\nsym}}$, see Eq. \ref{eq:rhoA}. Then (center) we use the invariance of $\ket{\Psi_{\nsym}}$ under the action of $\mathcal{G}$, Eq. \ref{eq:invstate}. Finally (right) we use the fact that $\mathcal{G}$ acts unitarily and therefore $(\hat{U}_{g})^{\dagger}\hat{U}_g = \hat{\mathbb{I}}$ for the sites that have been traced out.
 \label{fig:SymPsi}}
\end{figure}

\begin{figure}[t]
  \includegraphics[width=8.0cm]{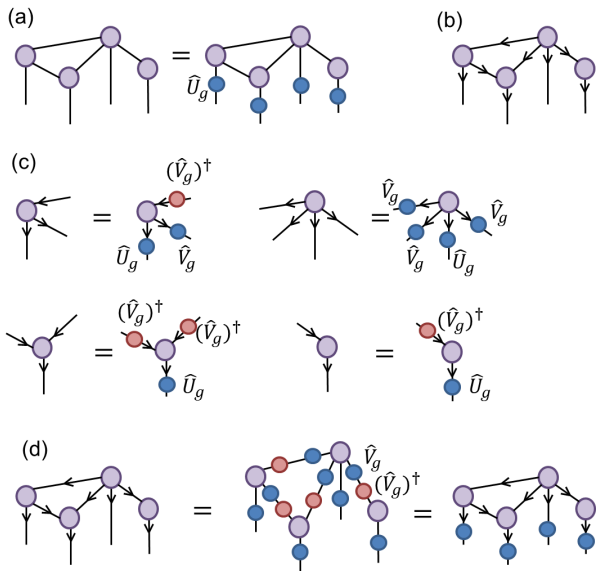}
\caption{
(a) Graphical representation of the global constraint fulfilled collectively by the tensors in a tensor network that represents a $\mathcal{G}$-symmetric state.
(b) Possible assignment of directions in the indices of a tensor network.
(c) Constraint fulfilled by each $\mathcal{G}$-symmetric tensor. Notice that the group acts with $\hat{V}_g$ on outgoing bond indices and with $(\hat{V}_g)^{\dagger}$ on incoming bond indices.
(d) Proof of lemma 2: (left) tensor network $\mathcal{N}_{\nsym}$ made of $\mathcal{G}$-symmetric tensors; (center) the same tensor network where we have used the invariance of each tensor under a symmetry transformation (here, each symmetry transformation implements the same group element $g\in \mathcal{G}$); (right) after using that $(\hat{V}_{g})^{\dagger} \hat{V}_g = \hat{\mathbb{I}}$ on each bond index, we are left with the action $\mathcal{G}$ only on the physical indices.
 \label{fig:SymTN}}
\end{figure}

\section{Tensor networks in the presence of a global on-site symmetry}
\label{sec:invtn}

In this section we review the use of $\mathcal{G}$-symmetric tensors in tensor networks.

\subsection{Symmetric many-body wave-function}

Let us consider a compact, completely reducible group $\mathcal{G}$, such as a discrete group (e.g. $\mathbb{Z}_2$ for particle number parity) or a Lie group (e.g. U(1) for particle number conservation or SO(3) for spin isotropy), acting on the vector space $\mathbb{V}^{\otimes L}$ of lattice $\mathcal{L}$ by means of an on-site unitary representation $(\hat{U}_g)^{\otimes L}$, where $\hat{U}_{g}:\mathbb{V} \rightarrow \mathbb{V}$ is a single-site unitary transformation and $g \in \mathcal{G}$ is an element of the group $\mathcal{G}$ \cite{unitary}.

We say that a many-body wave-function $\ket{\Psi_{\nsym}} \in \mathbb{V}^{\otimes L}$ is invariant under $\mathcal{G}$, or just $\mathcal{G}$-\textit{symmetric}, if
\begin{equation}
\ket{\Psi_{\nsym}} = (\hat{U}_{g})^{\otimes L} \ket{\Psi_{\nsym}}, ~~~\forall~g \in \mathcal{G}. \label{eq:invstate}
\end{equation}
In terms of the coefficients $\hat{\Psi}_{i_1i_2\ldots i_{L}}$ in Eq. \ref{eq:genstate}, this condition reads
\begin{equation}
\hat{\Psi}_{i'_1i'_2\ldots i'_{L}} = \sum_{i_1,i_2,\ldots,i_{L}}(\hat{U}_{g})_{i'_1i_1} (\hat{U}_{g})_{i'_2i_2} \ldots (\hat{U}_{g})_{i'_Li_L}\hat{\Psi}_{i_1i_2\ldots i_{L}}, \label{eq:invstatecomp}
\end{equation}
as represented diagrammatically in \fref{fig:SymPsi}(a). Let $\hat{\rho}^{\mathcal{A}}$ denote the reduced density matrix
\begin{equation}\label{eq:rhoA}
    \hat{\rho}^{\mathcal{A}} \equiv \tr_{\mathcal{B}} \left(\proj{\Psi_{\nsym}}\right),
\end{equation}
where $\mathcal{A}$ denotes a subset of $L_{\mathcal{A}}$ sites of lattice $\mathcal{L}$ and $\mathcal{B}$ the remaining sites.

\textbf{Lemma 1.} \textit{The reduced density matrix $\hat{\rho}^{\mathcal{A}}$ is also invariant under the action of the group, that is,}
\begin{align}
\hat{\rho}^{\mathcal{A}} &= \left( (\hat{U}_{g})^{\otimes L_{\mathcal{A}}}\right) \hat{\rho}^{\mathcal{A}} \left( (\hat{U}_{g}^{\dagger})^{\otimes L_{\mathcal{A}}} \right),
~~~\forall~g \in \mathcal{G}.
\label{eq:invrho}
\end{align}

This lemma follows immediately from Eqs. \ref{eq:invstate} and \ref{eq:rhoA} and the fact that $\hat{U}_g^{\dagger}\hat{U}_g = \hat{\mathbb{I}}$, see \fref{fig:SymPsi}(b).

\subsection{Tensor network made of symmetric tensors}

\fref{fig:SymTN}(a) shows a tensor network $\mathcal{N}$ for a $\mathcal{G}$-symmetric many-body wave-function $\ket{\Psi_{\nsym}}$. The global symmetry imposes collective constraints on the tensors. A possible way of satisfying these constraints is by using a tensor network $\mathcal{N}_{\nsym}$ made of $\mathcal{G}$-symmetric tensors. In order to do so, we first need to extend the action of the group $\mathcal{G}$ to the bond indices of the network. For this purpose we assign a direction to each bond index, see \fref{fig:SymTN}(b) and then define a unitary representation $\hat{V}_{g}:\mathbb{C}_{\chi} \rightarrow \mathbb{C}_{\chi}$, where $\mathbb{C}_{\chi}$ is a $\chi$-dimensional vector space associated to each bond index. [To keep the notation simple, in this section we assume that all bond indices are equivalent: they have the same bond dimension $\chi$ and $\mathcal{G}$ acts with the same unitary representation $\hat{V}_{g}$. However, all the results presented in this section generalize to the case where each bond index has its own bond dimension and unitary representation of $\mathcal{G}$].

Given a tensor $\hat{T}$ in the tensor network, we distinguish between outgoing bond indices, where the group acts with $\hat{V}_{g}$, and incoming bond indices, where the group acts with its adjoint $\hat{V}_g^{\dagger}$ (for consistency, we can think of the physical indices, where the group acts with $\hat{U}_g$, also as outgoing indices). We then say that tensor $\hat{T}$ is symmetric if it is invariant under the simultaneous action of $\mathcal{G}$ on all its indices. For instance, for the left-most tensor in the tensor network of \fref{fig:SymTN}(b), which has one physical index $i$ and two bond indices $j$ and $k$, where $i$ and $j$ are outgoing indices and $k$ is an incoming index, invariance under $\mathcal{G}$ is given by
\begin{equation}\label{eq:tensortransform}
\hat{T}_{i'j'k'} = \sum_{ijk} (\hat{U}_g)_{i'i} (\hat{V}_g)_{j'j} \hat{T}_{ijk} (\hat{V}^{\dagger}_g)_{kk'}, ~~~\forall g \in \mathcal{G},
\end{equation}
see \fref{fig:SymTN}(c). The following lemma is proved in \fref{fig:SymTN}(d).

\textbf{Lemma 2.} \textit{A tensor network $\mathcal{N}_{\nsym}$ made of symmetric tensors represents a symmetric many-body wave-function $\ket{\Psi_{\nsym}}$.}

As discussed in Refs. \cite{Singh10,Singh11,Singh12}, a $\mathcal{G}$-symmetric tensor can be regarded as a linear combination of \textit{intertwiners}, and as such be written in its $(P,Q)$-decomposition \cite{Singh10}, where $P$ are a collection of \textit{degeneracy} tensors, containing all the degrees of freedom not fixed by the symmetry, and $Q$ are \textit{structural} tensors (intertwiners, build from the Clebsh-Gordan coefficients) completely specified by the symmetry. The $(P,Q)$-decomposition of a $\mathcal{G}$-symmetric tensor offers a very compact representation, in which only the degeneracy tensors $P$ and some minimal information about quantum numbers need to be stored in memory. It also leads to very significant computational gains during tensor manipulations \cite{Singh10,Singh11,Singh12}.

Lemma 2 tells us that the use of $\mathcal{G}$-symmetric tensors is \textit{sufficient} to ensure the global symmetry of $\ket{\Psi_{\nsym}}$. However, it is not \textit{necessary}. Indeed, we had already mentioned that it is also possible to represent $\ket{\Psi_{\nsym}}$ by means of a tensor network $\mathcal{N}$ made of non-symmetric tensors, which are only \textit{collectively} constrained by the symmetry, \fref{fig:SymTN}(a). An immediate question is then whether the use of symmetric tensors, where each tensor is individually constrained by the symmetry, may result in a minimal bond dimension $\chi^{\nmin}_{\nsym}$ larger than the minimal bond dimension $\chi^{\nmin}$ achievable if the tensors are only collectively constrained by the symmetry. The next two sections give an answer to this question.

\begin{figure}[t]
  \includegraphics[width=8.0cm]{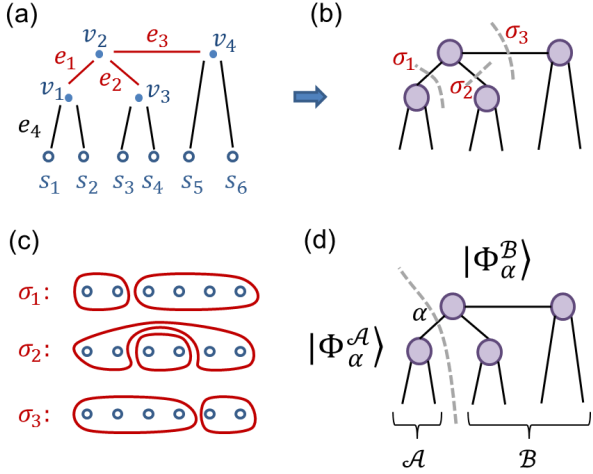}
\caption{
(a) Tree graph associated to a tree tensor network (TTN). It is made of a set $V$ of vertices, $V=\{v_1,v_2,\cdots, s_1, s_2, \cdots\}$ (where $\{s_1, s_2, \cdots\}$ denote the sites of lattice $\mathcal{L}$), and a set $E$ of edges,  $E=\{e_1,e_2,\cdots\}$, where each edge can be specified by a pair of vertices (for instance, $e_1 = (v_1,v_2)$ and $e_4 =(v_1,s_1)$). The defining property of a tree graph is the absence of loops.
(b) A TTN is obtained from a tree graph by replacing each vertex in $V$ (other than those corresponding to sites of $\mathcal{L}$) with a tensor, and each edge in $E$ with an index. Bond indices connect two tensors.
(c) Each bond index defines a bipartition $\sigma \in \Omega$. This TTN has three bond indices, and therefore $\Omega= \{\sigma_1, \sigma_2, \sigma_3\}$.
(d) A bond index $\alpha$ divides the TTN into two parts, and each part describes a set of $\chi$ states,  $\{\ket{\Phi^{\mathcal{A}}_{\alpha}} \}_{\alpha=1}^{\chi}$ and $\{\ket{\Phi^{\mathcal{B}}_{\alpha}} \}_{\alpha=1}^{\chi}$, where $\mathcal{A}$ and $\mathcal{B}$ are complementary subsets of sites of $\mathcal{L}$.
 \label{fig:TTN2}}
\end{figure}

\section{Tensor networks without loops}
\label{sec:ttn}

Let us consider a tensor network without loops, that is, a tree tensor network (TTN), which includes the matrix product state (MPS) with open boundary conditions (OBC) as a particular case, see \fref{fig:TTNMPS}.

\subsection{Tree tensor network}

An undirected graph is a pair $(V,E)$, where $V$ is a set of vertices and $E$ is a set of edges connecting the vertices. A tree is an undirected graph \textit{without} loops. Here we are interested in a tree with $L$ open edges or leaves, one for each site of lattice $\mathcal{L}$, see \fref{fig:TTN2}(a). We refer to any such tree as a \textit{tree on} $\mathcal{L}$. Let $\sigma$ denote a bipartition of lattice $\mathcal{L}$ into two subsets $\mathcal{A}$ and $\mathcal{B}$ of lattice sites, such that $\mathcal{A}\cap\mathcal{B} = \varnothing$ and  $\mathcal{A}\cup\mathcal{B} = \mathcal{L}$. Notice that each edge of a tree on $\mathcal{L}$ defines a different bipartition $\sigma$ of $\mathcal{L}$. Let then $\Omega$ be the set of bipartitions $\sigma$ of $\mathcal{L}$ induced by the tree, where we exclude the bipartitions induced by the tree where either part $\mathcal{A}$ or part $\mathcal{B}$ only contain one leaf, see \fref{fig:TTN2}(b). 

Given a tree $(V,E)$ on $\mathcal{L}$, a TTN is obtained by attaching a tensor to each vertex in $V$ and by then connecting the tensors with indices according to the pattern of edges in $E$, \fref{fig:TTN2}(a),(c). By construction, a bond index $\alpha$ of the TTN corresponds to a bipartition $\sigma \in \Omega$ of the lattice $\mathcal{L}$ into parts $\mathcal{A}$ and $\mathcal{B}$, which also divides the TTN into two parts, \fref{fig:TTN2}(d). Each of these two parts is in itself a TTN. Let $\chi$ denote the dimension of that bond index $\alpha$. Then one part of the tensor network describes a set of $\chi$ many-body states $\{\ket{\Phi^{\mathcal{A}}_{\alpha}} \}_{\alpha=1}^{\chi}$ for the sites in $\mathcal{A}$, and the other part of the tensor network describes a set of $\chi$ many-body states $\{\ket{\Phi^{\mathcal{B}}_{\alpha}} \}_{\alpha=1}^{\chi}$ for the sites in $\mathcal{B}$, see \fref{fig:TTN2}(d). Moreover, the many-body wave-function $\ket{\Psi}$ for the whole lattice $\mathcal{L}$ can be written as
\begin{equation}
\ket{\Psi} = \sum_{\alpha=1}^{\chi} \ket{\Phi_{\alpha}^{\mathcal{A}}} \ket{\Phi_{\alpha}^{\mathcal{B}}}, ~~~~~~~~~~\forall \sigma \in \Omega,
\end{equation}
from where it follows that the reduced density matrix $\hat{\rho}^{\mathcal{A}} \equiv \tr_{\mathcal{B}} \left( \proj{\Psi} \right)$ can be written as
\begin{equation}\label{eq:rhoAA}
\hat{\rho}^{\mathcal{A}} = \sum_{\alpha\alpha'=1}^{\chi} \braket{\Phi_{\alpha}^{\mathcal{B}}}{\Phi_{\alpha'}^{\mathcal{B}}} \ketbra{\Phi_{\alpha}^{\mathcal{A}}}{\Phi_{\alpha'}^{\mathcal{A}}}.
\end{equation}
Then this lemma follows simply from Eq. \ref{eq:rhoAA}.

\textbf{Lemma 3.} \textit{ For any bipartition $\sigma\in \Omega$, the rank $r(\hat{\rho}^{\mathcal{A}})$ of the reduced density matrix $\hat{\rho}^{\mathcal{A}}$ is less than or equal to the bond dimension $\chi$,}
\begin{equation}\label{eq:lemma3}
    r(\hat{\rho}^{\mathcal{A}}) \leq \chi.
\end{equation}

\subsection{Tree tensor network for a symmetric state}\label{ssec:proof}


Until the end of this section, each partition $\sigma\in \Omega$ has a bond index $\alpha^{\sigma}$ with bond dimension $\chi^{\sigma}$. That is, the different bond indices may have different bond dimension.

\begin{figure}[t]
  \includegraphics[width=8.0cm]{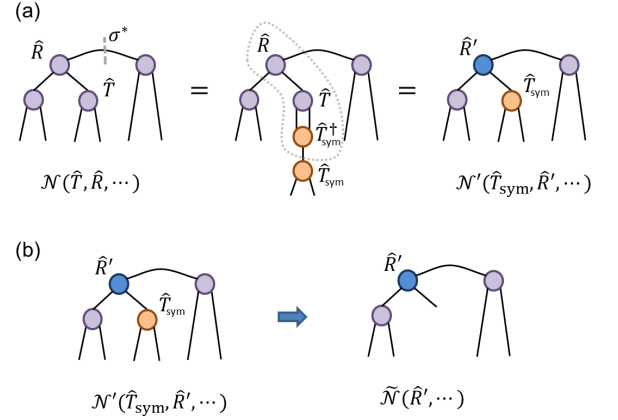}
\caption{
(a) Basic step of replacing a tensor $\hat{T}$ with a $\mathcal{G}$-symmetric tensor $\hat{T}_{\nsym}$ in a tensor network $\mathcal{N}$.
(b) Tensor network $\mathcal{N}'$, containing the same number of tensors as $\mathcal{N}$, and tensor network $\tilde{\mathcal{N}}$, which contains one tensor less.
\label{fig:SymTTN}}
\end{figure}

\textbf{Theorem.} Let $\mathcal{N}$ denote a TTN for $\ket{\Psi_{\nsym}} \in \mathbb{V}^{\otimes L}$, according to a tree with bipartitions $\sigma \in \Omega$ and with bond dimensions $\{\chi^{\sigma}\}_{\sigma \in \Omega}$. Then there exist a TTN $\mathcal{N}_{\nsym}$ for $\ket{\Psi_{\nsym}}$ that is made of symmetric tensors and such that its bond dimensions $\{\chi^{\sigma}_{\nsym}\}_{\sigma \in \Omega}$ fulfill
\begin{equation}\label{eq:Teorem}
    \chi^{\sigma}_{\nsym} \leq \chi^{\sigma},~~~~~~~\forall \sigma \in \Omega.
\end{equation}

\textbf{Proof.} We will prove this result constructively, producing the TTN $\mathcal{N}_{\nsym}$ by sequentially replacing each tensor in $\mathcal{N}$ with a symmetric tensor. For simplicity, we assume that $\mathcal{N}$ is made of trivalent tensors, although the proof can be easily generalized to the case where the tensors have more than three indices \cite{nonTrivalentTTN}. First, we (arbitrarily) choose one bipartition $\sigma^{*}$ of the TTN, \fref{fig:SymTTN}, which will be treated differently. 

\begin{figure}[t]
  \includegraphics[width=8cm]{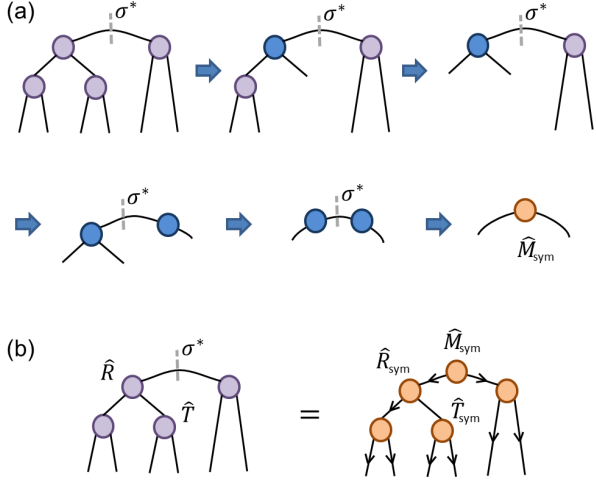}
\caption{
(a) Sequence of tensor networks with non-symmetric tensors, obtained by iteratively using the basic step (described in the text), where at the end of each step a $\mathcal{G}$-symmetric tensor is removed.
(b) A tensor network $\mathcal{N}_{\nsym}$ made of $\mathcal{G}$-symmetric tensors is obtained by connecting all the $\mathcal{G}$-symmetric tensors resulting from iterating the basic step, and the matrix $\hat{M}_{\nsym}$.
\label{fig:Sequence}}
\end{figure}

\textit{Basic step-.} Let $\sigma \in \Omega$ denote a bipartition of lattice $\mathcal{L}$ into a part $\mathcal{A}$ containing two sites $s$ and $s'$ and a part $\mathcal{B}$ containing the remaining sites. Let $\hat{T}$ be the tensor in the TTN $\mathcal{N}$ that connects the indices $i$ and $i'$ for sites $s$ and $s'$ with the bond index $\alpha$ for partition $\sigma$. (That is, $\sigma$ corresponds to a bipartition of the tensor network into a single tensor $\hat{T}$ and the remaining tensors.) We can always regard $\hat{T}$ as a linear map
\begin{equation}\label{eq:T}
    \hat{T} : \mathbb{V}^{\sigma} \rightarrow \mathbb{V}_{\nsym}^{s}\otimes \mathbb{V}_{\nsym}^{s'},
\end{equation}
where $\mathbb{V}^{\sigma}$ denotes a vector space of dimension $\chi^{\sigma}$ for the bond index $\alpha$, and $\mathbb{V}_{\nsym}^{s}$ and $\mathbb{V}_{\nsym}^{s'}$ denote the vector spaces of sites $s$ and $s'$, for which we use the subscript `sym' to emphasize that the symmetry group $\mathcal{G}$ acts with unitary representations
\begin{equation}\label{eq:UV}
    \hat{U}^{(s)}_g:\mathbb{V}_{\nsym}^{s} \rightarrow \mathbb{V}_{\nsym}^{s}, ~~~~~
    \hat{U}^{(s')}_g: \mathbb{V}_{\nsym}^{s'} \rightarrow \mathbb{V}_{\nsym}^{s'}.
\end{equation}
Let $\hat{\rho}^{\mathcal{A}}$ be the reduced density matrix for part $\mathcal{A}$,
\begin{equation}\label{eq:rhoAAA}
    \hat{\rho}^{\mathcal{A}}: \mathbb{V}_{\nsym}^{s} \otimes \mathbb{V}_{\nsym}^{s'} \rightarrow \mathbb{V}_{\nsym}^{s} \otimes \mathbb{V}_{\nsym}^{s'}.
\end{equation}
According to lemma 1, $\hat{\rho}^{\mathcal{A}}$ is $\mathcal{G}$-symmetric. Let us consider its eigenvalue decomposition,
\begin{equation}\label{eq:eigrho}
\hat{\rho}^{\mathcal{A}} = \hat{Q}\hat{D}\hat{Q}^\dagger,
\end{equation}
where $\hat{Q}$ is a unitary matrix ($\hat{Q} \hat{Q}^{\dagger} = \hat{Q}^{\dagger} \hat{Q}= \hat{\mathbb{I}}$) and $\hat{D}$ is a diagonal matrix with the (non-negative) eigenvalues of $\hat{\rho}^{\mathcal{A}}$ in its diagonal. Notice that $\hat{Q}$ is a $\mathcal{G}$-symmetric matrix \cite{invariantSVD}. In addition, matrix $\hat{D}$ only contains some number $\chi_{\nsym}^{\sigma}$ of non-zero eigenvalues, where (by lemma 3) $\chi_{\nsym}^{\sigma} \leq \chi^{\sigma}$. Let $\hat{P}$ be a projector onto the non-zero eigenvalue subspace $\mathbb{V}_{\nsym}$ of $\hat{D}$. Then $\hat{T}_{\nsym} \equiv \hat{Q}\hat{P}$ is an isometry,
\begin{equation}\label{eq:Tsym}
    \hat{T}_{\nsym}: \mathbb{V}_{\nsym} \rightarrow \mathbb{V}^{s}\otimes \mathbb{V}^{s'}, ~~~~~(\hat{T}_{\nsym})^{\dagger}\hat{T}_{\nsym} = \hat{\mathbb{I}}.
\end{equation}
Further, let us define a unitary representation $\hat{V}^{(\sigma)}_g$ of $\mathcal{G}$ by
\begin{equation}\label{eq:Vg}
    \hat{V}^{(\sigma)}_g \equiv (\hat{T}_{\nsym})^{\dagger} \left(\hat{U}^{(s)}_g \otimes \hat{U}^{(s')}_g \right) (\hat{T}_{\nsym}), ~~~~\forall g\in \mathcal{G}.
\end{equation}
Notice that
\begin{equation}\label{eq:TsymPsi}
    \hat{T}_{\nsym}(\hat{T}_{\nsym})^{\dagger}\ket{\Psi_{\nsym}} = \ket{\Psi_{\nsym}},
\end{equation}
since the projector $\hat{T}_{\nsym}(\hat{T}_{\nsym})^{\dagger}$ preserves the support of $\hat{\rho}^{\mathcal{A}}$. Consequently, we can obtain a new TTN $\mathcal{N}'$ for $\ket{\Psi_{\nsym}}$ from the old TTN $\mathcal{N}$
\begin{equation}\label{eq:NN}
    \mathcal{N}(\hat{T}, \hat{R}, \hat{S}, \cdots) \Longrightarrow \mathcal{N}'(\hat{T}_{\nsym}, \hat{R}', \hat{S}, \cdots),
\end{equation}
by replacing tensor $\hat{T}$ with tensor $\hat{T}_{\nsym}$, and the tensor $\hat{R}$ that is connected to $\hat{T}$ with a new tensor $\hat{R}' = (\hat{T}_{\nsym})^{\dagger} \hat{T} \hat{R}$, \fref{fig:SymTTN}(a). Since the matrix $(\hat{T}_{\nsym})^{\dagger} \hat{T}$ maps $\mathbb{V}^{\sigma}$ into $\mathbb{V}^{\sigma}_{\nsym}$,
\begin{equation}\label{eq:TT}
(\hat{T}_{\nsym})^{\dagger} \hat{T}: \mathbb{V}^{\sigma} \rightarrow \mathbb{V}^{\sigma}_{\nsym},
\end{equation}
it follows that the index $\alpha$ of $\hat{R}$ becomes an index $\alpha'$ in $\hat{R}'$ with a $\chi_{\nsym}^{\sigma}$-dimensional vector space $\mathbb{V}^{\sigma}_{\nsym}$ on which the symmetry group $\mathcal{G}$ acts with the unitary representation $\hat{V}_g$.
In summary, then, we have been able to replace one tensor $\hat{T}$ with a $\mathcal{G}$-symmetric tensor $\hat{T}_{\nsym}$, without increasing the corresponding bond dimension.

\textit{Iteration-.} Next, consider the TTN $\tilde{\mathcal{N}}(\hat{R}',\hat{S}, \cdots)$ obtained by removing tensor $\hat{T}_{\nsym}$ from $\mathcal{N}'$, \fref{fig:SymTTN}(b). By construction, this new TTN describes a symmetric state $\ket{\tilde{\Psi}_{\nsym}}$ on a lattice $\tilde{\mathcal{L}}$ which is obtained from lattice $\mathcal{L}$ by replacing sites $s$ and $s'$ with an effective site $\tilde{s}$. This effective site $\tilde{s}$ has vector space $\mathbb{V}^{\sigma}_{\nsym}$ and unitary representation $\hat{V}^{(\sigma)}_g$.
We can therefore repeat the basic step above for $\tilde{\mathcal{N}}$, where we choose another suitable bipartition $\tilde{\sigma} \in \Omega$, and iterate for all $L-3$ bipartitions $\sigma \in \Omega$, including bipartition $\sigma^{*}$, which is chosen twice, see \fref{fig:Sequence}.

\textit{Final step-.} After $L-2$ iterations of the basic step, \fref{fig:Sequence}(a), we are left with one tensor that describes a $\mathcal{G}$-symmetric state of two effective sites, corresponding to bipartition $\sigma^{*}$. This tensor is therefore a $\mathcal{G}$-symmetric matrix $\hat{M}_{\nsym}$.

The tensor network $\mathcal{N}_{\nsym}$ for $\ket{\Psi_{\nsym}} \in \mathbb{V}^{\otimes L}$ is then obtained by collecting the $L-2$ $\mathcal{G}$-symmetric, isometric tensors, such as the one in Eq. \ref{eq:Tsym}, and the $\mathcal{G}$-symmetric matrix $\hat{M}_{\nsym}$, \fref{fig:Sequence}(b).\markend

\section{Tensor networks with loops}
\label{sec:tnloops}

In this section we build two explicit examples of tensor networks that contain loops and for which the replacement of their tensors with $\mathcal{G}$-symmetric tensors necessarily leads to larger bond dimensions.

\subsection{Matrix product state (MPS) with periodic boundary conditions (PBC)} \label{sec:mpspbc}

For our first example, let us consider a lattice $\mathcal{L}$ made of $L=3$ sites, each one with vector space $\mathbb{V}\cong \mathbb{C}_3$,  and the state
\begin{equation}\label{eq:Psi}
  \ket{\Psi} \equiv \frac{1}{\sqrt{3}}\left( \ket{0 0 0} + \ket{1 1 1} + \ket{2 2 2}\right) \in (\mathbb{C}_{3})^{\otimes 3},
\end{equation}
which is invariant under a global $\mathcal{G} = \mathbb{Z}_3$ symmetry, generated by $\hat{U}\otimes \hat{U} \otimes \hat{U}$, where
\begin{equation}\label{eq:U}
  \hat{U} = \left(
        \begin{array}{ccc}
          0 & 1 & 0 \\
          0 & 0 & 1 \\
          1 & 0 & 0 \\
        \end{array}
      \right), ~~~~~~~~~ \hat{U}^3 = I.
\end{equation}

\textbf{Lemma 4.} \textit{The state $\ket{\Psi}$ in \eref{eq:Psi} can be represented with a MPS with PBC where the dimension of each of the three bond indices is $\chi^{\min} = 2$.}

Indeed, let us define the sets of $2\times 2$ MPS matrices $\{\hat{A}^{i}, \hat{B}^{i}, \hat{C}^{i}\}$, where $i=0,1,2$, with all the matrix coefficients equal to zero except for the following ones:
\begin{eqnarray}
  \sqrt{3}\hat{A}^{0}_{10} &=& \hat{B}^{0}_{01} = \hat{C}^{0}_{11} = 1\\
  \sqrt{3}\hat{A}^{1}_{01} &=& \hat{B}^{1}_{11} = \hat{C}^{1}_{10} = 1\\
  \sqrt{3}\hat{A}^{2}_{11} &=& \hat{B}^{2}_{10} = \hat{C}^{2}_{01} = 1.
\end{eqnarray}
Then we have that, indeed,
\begin{eqnarray}\label{eq:MPS}
  &&\sum_{i,j,k=0}^{2} \tr (\hat{A}^{i}\hat{B}^{j}\hat{C}^{k})\ket{ijk} \\
  &=& \left(\hat{A}^{0}_{10} \hat{B}^{0}_{01} \hat{C}^{0}_{11}\right) \ket{000}
  + \left(\hat{A}^{1}_{01} \hat{B}^{1}_{11} \hat{C}^{1}_{10}\right) \ket{111} \\
  &+& \left(\hat{A}^{2}_{11} \hat{B}^{2}_{10} \hat{C}^{2}_{01}\right) \ket{222}
  = \frac{1}{\sqrt{3}} \sum_{i=0}^2 \ket{iii}.
\end{eqnarray}

This specific MPS with PBC has the triplet of bond dimensions $(2,2,2)$. Notice that if the bond dimension $\chi$ was less than $2$ for one index, then we would have an MPS with OBC. It is then easy to see that the remaining bond dimensions must be at least $\chi=3$, that is, we would have bond dimensions $(1,3,3)$.

\begin{figure}[t]
  \includegraphics[width=5cm]{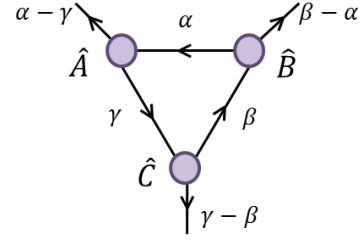}\\
  \caption{
MPS with PBC made of three $\mathbb{Z}_3$-symmetric tensors $\{\hat{A}, \hat{B}, \hat{C}\}$. Each index of each tensor is labeled by irreducible representations of $\mathbb{Z}_3$. For instance, tensor $\hat{A}^{\mu}_{\gamma\alpha}$ has indices with irreducible representations $\gamma$, $\alpha$ and $\mu = \alpha-\gamma|$ (modulo 3), where we have already taken into account the constraints imposed by the symmetry, namely that the sum (modulo 3) of incoming charges be equal to the sum (modulo 3) of outgoing charges (see e.g. \cite{Singh11}).
  }\label{fig:MPSPBC}
\end{figure}

Let us now try to obtain an MPS with PBC with bond dimensions $(2,2,2)$ using $\mathbb{Z}_3$-symmetric tensors. In this case, each index of a tensor labels irreducible representations (irreps) of the group $\mathbb{Z}_3$. All irreps are one-dimensional, since $\mathbb{Z}_3$ is an Abelian group. There are three inequivalent irreps, namely $1,e^{i2\pi/3}$ and $e^{i4\pi/3}$. Let us consider first the open indices, corresponding to states of each lattice site. On each site, the irreps are given by the states
\begin{eqnarray}
  \ket{\tilde{0}} &\equiv& (\ket{0} + ~~~\ket{1} + ~~~\ket{2})/\sqrt{3}, \\
  \ket{\tilde{1}} &\equiv& (\ket{0} + ~w\ket{1} + w^2\ket{2})/\sqrt{3}, \\
  \ket{\tilde{2}} &\equiv& (\ket{0} + w^2\ket{1} + ~w\ket{2})/\sqrt{3},
  \end{eqnarray}
where $w \equiv e^{i2\pi/3}$. Inverting this relation,
\begin{eqnarray}
  \ket{0} &\equiv& (\ket{\tilde{0}} + ~~~\ket{\tilde{1}} + ~~~\ket{\tilde{2}})/\sqrt{3}, \\
  \ket{1} &\equiv& (\ket{\tilde{0}} + w^2\ket{\tilde{1}} + ~w\ket{\tilde{2}})/\sqrt{3}, \\
  \ket{2} &\equiv& (\ket{\tilde{0}} + ~w\ket{\tilde{1}} + w^2\ket{\tilde{2}})/\sqrt{3},
\end{eqnarray}
we can express $\ket{\Psi}$ as
\begin{eqnarray}
  \ket{\Psi} = &&~~(~~\ket{\tilde{0}\tilde{0}\tilde{0}} + \ket{\tilde{0}\tilde{1}\tilde{2}} + \ket{\tilde{0}\tilde{2}\tilde{1}} \nonumber \\
  &&+ ~~~ \ket{\tilde{1}\tilde{0}\tilde{2}} + \ket{\tilde{1}\tilde{1}\tilde{1}} + \ket{\tilde{1}\tilde{2}\tilde{0}} \nonumber \\
  &&+ ~~~ \ket{\tilde{2}\tilde{0}\tilde{1}} + \ket{\tilde{2}\tilde{1}\tilde{0}} + \ket{\tilde{2}\tilde{2}\tilde{2}} ~~)/3.\label{eq:z3basis}
\end{eqnarray}

\textbf{Lemma 5.} \textit{If $\ket{\Psi}$ in \eref{eq:Psi} is represented by an MPS with PBC made of $\mathbb{Z}_3$-symmetric tensors, then at least one of the bond dimensions must be larger than or equal to 3.}

Indeed, let us suppose to the contrary that each bond index is equal to 2. That is, two different charges appear on each bond index. As shown in \fref{fig:MPSPBC}, any choice of charges on the bond indices uniquely determines the charges on the open indices, if the tensors are to be $\mathbb{Z}_3$-symmetric. Therefore, with only 2 charges on each bond index at most 8 triples of charges $(\alpha-\gamma, \beta-\alpha, \gamma-\beta)$ can be obtained on the open indices. However, 9 different triples of charges appear in \eref{eq:z3basis}. Thus, to allow for all 9 triples of charges on the open indices, at least one bond dimension must be 3.

We have then seen that the presence of a closed loop in an MPS with PBC implies that it is not always possible to replace non-symmetric tensors (with a triple of bond dimensions (2,2,2)) with symmetric tensors without increasing at least one of the bond dimensions. That is, $\chi^{\nmin}_{\nsym} > \chi^{\nmin}$ when representing the state $\ket{\Psi}$ in \eref{eq:Psi} with a PBC MPS.

\begin{figure}[t]
  \includegraphics[width=8cm]{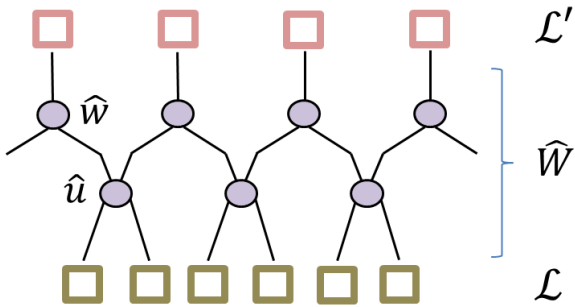}
  \caption{
\label{fig:MERA}
Coarse-graining transformation $\hat{W}$ made of disentanglers $\hat{u}$ and isometries $\hat{w}$. Each square at the bottom of the figure represents one site of lattice $\mathcal{L}$, with vector space $\mathbb{V} \cong (\mathbb{C}_2)_{\vartriangleleft} \otimes (\mathbb{C}_2)_{\vartriangleright}$, whereas each square at the top represents one site of the coarse-grained lattice $\mathcal{L}'$, with vector space $\mathbb{V}'$.}
\end{figure}

\subsection{Multi-scale entanglement renormalization ansatz (MERA)} \label{sec:mera}
 
For our second example, let us consider a lattice $\mathcal{L}$ in $D=1$ spatial dimensions made of $L$ sites, where each site has a four-dimensional vector space $\mathbb{V} \cong (\mathbb{C}_2)_{\vartriangleleft} \otimes (\mathbb{C}_2)_{\vartriangleright}$ that decomposes as the tensor product of two qubits, that we call \textit{left} qubit $(\mathbb{C}_2)_{\vartriangleleft}$ and \textit{right} qubit $(\mathbb{C}_2)_{\vartriangleright}$.
Let \textit{time reversal} act anti-unitarily on each qubit by means of $\hat{\mathcal{T}} \equiv i\hat{\sigma}_{y} \hat{\mathcal{K}}$, where $\hat{\sigma}_{y}$ is the Pauli matrix
\begin{equation}\label{eq:sigmay}
\hat{\sigma}_{y}\equiv \left(\begin{array}{cc} 0 & -i \\ i & 0 \end{array} \right),
\end{equation}
and $\hat{\mathcal{K}}$ denotes complex conjugation \cite{TimeRev}. Applying time reversal twice on a qubit amounts to applying the operator $(\mathcal{T})^2 = -\hat{\mathbb{I}}$. This implies that on each site of lattice $\mathcal{L}$, which is made of two qubits, time reversal squares to the identity operator, $(-\hat{\mathbb{I}})_{\vartriangleleft} \otimes (-\hat{\mathbb{I}})_{\vartriangleright} = \hat{\mathbb{I}}_{\mathbb{V}}$, defining a $\mathbb{Z}_2$ group of transformations. Following \cite{Chen2, Chen3, Chen4}, we denote this group of transformations by $\mathbb{Z}_2^T$, where the superscript $T$ emphasizes that it implements time reversal.

Still following \cite{Chen2, Chen3, Chen4}, on the above lattice, we consider a state $\ket{\Psi}\in \mathbb{V}^{\otimes L}$ that is the product of singlet states $\ket{\psi^{-}} \equiv \frac{1}{\sqrt{2}}(\ket{01}-\ket{10})$,
\begin{eqnarray}\label{eq:PsiMERA}
    \ket{\Psi} &\equiv& \bigotimes_{s\in \mathcal{L}} \frac{1}{\sqrt{2}}\left( \ket{0_{s,\vartriangleright} 1_{s+1,\vartriangleleft}} - \ket{1_{s,\vartriangleright} 0_{s+1,\vartriangleleft}} \right)\\
    &=& \bigotimes_{s\in \mathcal{L}} \ket{\psi^{-}}_{s,\vartriangleright;s+1,\vartriangleleft}
\end{eqnarray}
where each singlet state is an entangled state of the right qubit $(\mathbb{C}_2)_{\vartriangleright}$ of one site $s$ with the left qubit $(\mathbb{C}_2)_{\vartriangleleft}$ of the next site $s+1$, see \fref{fig:MERA}(a). This state is invariant under the action of time reversal $\mathbb{Z}_2^T$, as can be checked by noticing that each singlet state $\ket{\psi^-}$ fulfills
\begin{equation}\label{eq:Tsinglet}
    (i\sigma_y)\otimes(i\sigma_y)  \ket{\psi^{-}} = \ket{\psi^{-}}.
\end{equation}

Let us consider a MERA representation for the state $\ket{\Psi}$ in Eq. \ref{eq:PsiMERA}. Specifically, we focus on one layer $\hat{W}$ of the MERA, made of disentanglers $\hat{u}$ and isometries $\hat{w}$,
\begin{equation}\label{eq:MERA1}
    \hat{u}: \tilde{\mathbb{V}}\otimes \tilde{\mathbb{V}} \rightarrow \mathbb{V} \otimes \mathbb{V}, ~~~~~~ \hat{w}: \mathbb{V}' \rightarrow \tilde{\mathbb{V}} \otimes \tilde{\mathbb{V}},
\end{equation}
subject to isometric constraints
\begin{equation}\label{eq:uw}
    (\hat{u})^{\dagger} \hat{u} = \hat{\mathbb{I}}_{\tilde{\mathbb{V}}} \otimes \hat{\mathbb{I}}_{\tilde{\mathbb{V}}}, ~~~~~~ (\hat{w})^{\dagger} \hat{w}= \hat{\mathbb{I}}_{\mathbb{V}'},
\end{equation}
which implements a coarse-graining transformation from lattice $\mathcal{L}$ to a new lattice $\mathcal{L}'$ made of $L'=L/2$ sites, see \fref{fig:MERA}. In Eq. \ref{eq:MERA1}, $\tilde{\mathbb{V}}$ denotes the (possibly reduced) vector space of one site of $\mathcal{L}$ after the action of a disentangler $\hat{u}$, whereas $\mathbb{V}'$ is the vector space of one effective site in $\mathcal{L}'$. For each even $s$, first $\hat{u}^{\dagger}$ acts on pairs of sites $(s,s+1)$ of $\mathcal{L}$, and then $\hat{w}^{\dagger}$ acts on pairs of sites $(s-1,s)$, to produce a site of $\mathcal{L}'$. Overall, $\hat{W}^{\dagger}$ takes two adjacent sites of lattice $\mathcal{L}$, with combined vector space $\mathbb{V}\otimes\mathbb{V}$, into a single effective site of $\mathcal{L}'$, with vector space $\mathbb{V'}$. The coarse-graining transformation $\hat{W}^{\dagger}$ thus maps the state $\ket{\Psi}\in \mathbb{V}^{\otimes L}$ of the lattice $\mathcal {L}$ into a state
\begin{equation}\label{eq:Psiprime}
    \ket{\Psi'} \equiv \hat{W}^{\dagger} \ket{\Psi},
\end{equation}
of the coarse-grained lattice $\mathcal{L}'$. Our goal is to represent the state $\ket{\Psi}$ as a tensor network by writing it in terms of $\hat{W}$ and $\ket{\Psi'}$,
\begin{equation}\label{eq:PsiPsiprime}
    \ket{\Psi} = \hat{W}\ket{\Psi'}.
\end{equation}
This is actually just one step in building the MERA, which in general will concatenate several coarse-graining transformations, but it turns out to be sufficient for our purposes. 

Next we discuss whether there exist disentanglers $\hat{u}$ and isometries $\hat{w}$ such that the coarse-grained state $\ket{\Psi}'$ is a product state, which corresponds to a vector space $\mathbb{V}'$ of dimension $\chi=1$, or must necessarily remain entangled, which corresponds to a vector space $\mathbb{V}'$ of dimension $\chi>1$. Notice that here $\chi$ is the bond dimension of the upper index of the isometry $\hat{w}$.

\textbf{Lemma 6.} \textit{If the disentanglers $\hat{u}$ and isometries $\hat{w}$ are not restricted by time-reversal $\mathbb{Z}^T_2$ symmetry, then there exist tensors $(\hat{u},\hat{w})$ such that $\chi=1$.}

\textbf{Lemma 7.} \textit{If the disentanglers $\hat{u}_{\nsym}$ and isometries $\hat{w}_{\nsym}$ are restricted to be (time-reversal) $\mathbb{Z}^T_2$-symmetric, there do not exist tensors $(\hat{u}_{\nsym},\hat{w}_{\nsym})$ such that $\chi=1$.}

In other words, when representing $\ket{\Psi}$ as a MERA, the use of $\mathbb{Z}_2$-symmetric tensors requires a larger bond dimension, $\chi^{\nmin}_{\nsym} > \chi^{\nmin}$.

\begin{figure}[t]
  \includegraphics[width=8cm]{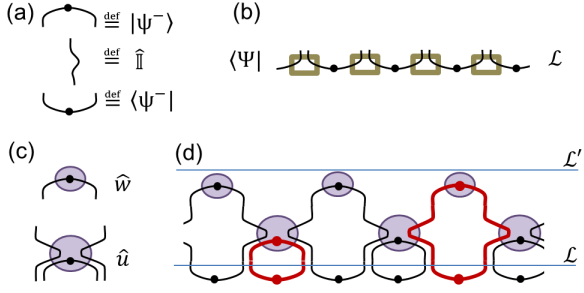}
  \caption{
(a) Graphical notation for $\ket{\psi^{-}}$, the identity operator on a qubit $\mathbb{C}_2$, and $\bra{\psi^{-}}$. In particular, a line represents the identity on a two-dimensional vector space.
(b) Graphical notation for $\bra{\Psi}$, where $\ket{\Psi}$ is the state in Eq. \ref{eq:PsiMERA}. 
(c) Isometry $\hat{w}$ and isometry $\hat{u}$ in Eqs. \ref{eq:uchi1}-\ref{eq:wchi1}. The lower legs of $\hat{u}$ are represented by a double line, representing the identity in the vector space $\mathbb{V} \cong (\mathbb{C}_2)_{\vartriangleleft} \otimes (\mathbb{C}_2)_{\vartriangleright}$, whereas the upper legs are represented by a single line, representing the identity in the vector space $\tilde{\mathbb{V}} \cong \mathbb{C}_2$. The isometry $\hat{w}$ has no upper leg, since the effective sites of $\mathcal{L}'$ have vector space $\mathbb{V}' \cong \mathbb{C}$.
(d) Under coarse-graining by $\hat{W}^{\dagger}$, the state $\ket{\Psi}$ becomes a product state. The figure shows $\bra{\Psi}\hat{W}$ or, equivalently, $\bra{\Psi'}$, where $\ket{\Psi'} = \ket{0}^{\otimes L/2}$. In red, we can follow the fate of two singlets $\ket{\psi^{-}}$ under coarse-graining. One of them (left) is absorbed  by a disentangler, whereas the other (right) is absorbed by an isometry.
\label{fig:MERAproduct}
}
\end{figure}

The proofs of lemma 6 and lemma 7 are closely related to recent results concerning symmetry protected phases of quantum matter in $D=1$ spatial dimension \cite{Pollmann, Chen1, Schuch11, Chen2}. For lemma 6 we can build the following disentangler $\hat{u}$ and isometry $\hat{w}$, Fig. \ref{fig:MERAproduct}(c),
\begin{eqnarray}\label{eq:uchi1}
    \hat{u} &\equiv& \hat{\mathbb{I}}_{s,\vartriangleleft} \otimes \left(\ket{\psi^-}_{s,\vartriangleright;s+1,\vartriangleleft} \right) \otimes \hat{\mathbb{I}}_{s+1,\vartriangleright},\\
    \label{eq:wchi1}
    \hat{w} &\equiv&  \left(  \ket{\psi^-}_{s-1,\vartriangleright;s,\vartriangleleft} \right) \bra{0}.
\end{eqnarray}
In this case, $\tilde{\mathbb{V}} \cong \mathbb{C}_2$, $\mathbb{V}' \cong \mathbb{C}$. When acting on $\ket{\Psi}$ as
\begin{equation}\label{eq:disentanglers}
    \ket{\Psi'} \equiv \left( \bigotimes_{\mbox{\tiny even }s} \hat{w}_{s-1,s}^{\dagger} \right) \left( \bigotimes_{\mbox{\tiny even }s} \hat{u}_{s,s+1}^{\dagger} \right) \ket{\Psi}
\end{equation}
the disentanglers eliminate every second singlet $\ket{\psi^{-}}$ in $\ket{\Psi}$, whereas the isometries eliminate the remaining singlets, Fig. \ref{fig:MERAproduct}(d). The resulting state is the product state
\begin{equation}\label{eq:prod}
    \ket{\Psi'} = \ket{0}^{\otimes L/2},
\end{equation}
which requires $\chi^{\nmin}=1$. This proves lemma 6. It turns out that by using $\mathbb{Z}_2^T$-symmetric local unitary transformations, it is not possible to transform $\ket{\Psi}$ into a product state. Since both $\hat{u}_{\nsym}$ and $\hat{w}_{\nsym}$ can always be completed into $\mathbb{Z}_2$-symmetric unitary transformations, the results of Refs. \cite{Pollmann, Chen1, Schuch11, Chen2} imply that by means of $\mathbb{Z}_2^T$-symmetric disentanglers and isometries it is not possible to transform $\ket{\Psi}$ into a product state. Therefore, we have that $\chi_{\nsym} > 1$, proving lemma 7.

\begin{figure}[t]
  \includegraphics[width=8cm]{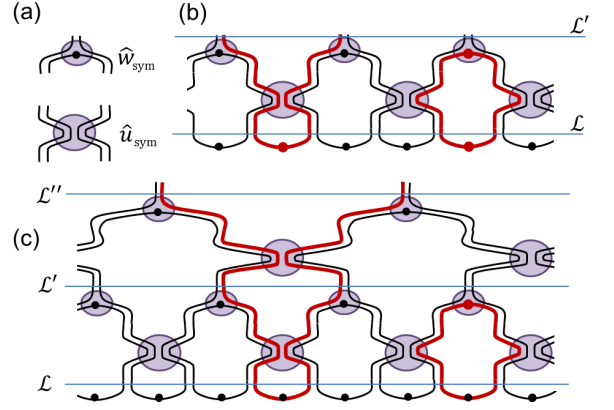}
  \caption{
(a) $\mathbb{Z}^T_2$-symmetric isometry $\hat{w}_{\nsym}$ and (trivial) disentangler $\hat{u}_{\nsym}$ in Eqs. \ref{eq:usym}-\ref{eq:wsym}. Each upper and lower leg of these tensors is represented by a double line, corresponding to the identity operator on a vector space $\mathbb{C}_2 \otimes \mathbb{C}_2$.
(b) Under coarse-graining by $\hat{W}_{\nsym}^{\dagger}$, the state $\ket{\Psi}$ is mapped into a state with the same local structure. The figure shows $\bra{\Psi}\hat{W}_{\nsym}$, which is equal to $\bra{\Psi}$ on the coarse-grained lattice $\mathcal{L}'$ with $L/2$ sites. In red, we can follow the fate of two singlets $\ket{\psi^{-}}$ under coarse-graining. One of them (left) propagates from $\mathcal{L}$ to $\mathcal{L}'$, whereas the other (right) is absorbed by an isometry.
(c) Concatenation of two steps of the coarse-graining transformation, to emphasize that the state $\ket{\Psi}$ is indeed a fixed-point of the $\mathbb{Z}^T_2$-symmetry protected coarse-graining transformation. At each coarse-graining step, half of the singlet states $\ket{\psi^-}$ are absorbed by the isometries.
\label{fig:MERAZ2}
}
\end{figure}

In preparation for the next section, let us introduce a second coarse-graining transformation $\hat{W}_{\nsym}$, made of (time-reversal) $\mathbb{Z}^T_2$-symmetric tensors, that has the state $\ket{\Psi}$ in \eref{eq:PsiMERA} as a fixed point. It consists of the following (trivial) disentanglers $\hat{u}_{\nsym}$ and isometries $\hat{w}_{\nsym}$ 
\begin{eqnarray}
\label{eq:usym}
    \hat{u}_{\nsym} &\equiv& \hat{\mathbb{I}}_s\otimes \hat{\mathbb{I}}_{s+1},\\
\label{eq:wsym}
    \hat{w}_{\nsym} &\equiv& \left(\hat{\mathbb{I}}_{s-1,\vartriangleleft} \otimes  \ket{\psi^-}_{s-1,\vartriangleright;s,\vartriangleleft} \otimes \hat{\mathbb{I}}_{s,\vartriangleright}\right),
\end{eqnarray}
Fig. \ref{fig:MERAZ2}(a). In this case $\tilde{\mathbb{V}} \cong (\mathbb{C}_2)_{\vartriangleleft} \otimes (\mathbb{C}_2)_{\vartriangleright}$, whereas $\mathbb{V}'$ for an effective site is obtained by joining left and right qubits from two different sites of $\mathcal{L}$, $\mathbb{V}' \cong (\mathbb{C}_2)_{s,\vartriangleleft} \otimes (\mathbb{C}_2)_{s+1,\vartriangleright}$, see Fig. \ref{fig:MERAZ2}(b). The coarse-grained state $\ket{\Psi'} = \hat{W}_{\nsym}^{\dagger} \ket{\Psi}$ is locally identical to $\ket{\Psi}$, and therefore $\ket{\Psi}$ is indeed a fixed-point of the coarse-graining transformation $\hat{W}_{\nsym}$.

We say that $\ket{\Psi}$ is a fixed-point of a symmetry protected (with symmetry $\mathbb{Z}_2^T$) renormalization group flow, in that we have enforced the $\mathbb{Z}_2^T$ symmetry in the coarse-graining tensors $\hat{u}_{\nsym}$ and $\hat{w}_{\nsym}$. If we do not enforce this symmetry on the tensors, then we have seen above that the state $\ket{\Psi}$ can be coarse-grained into a product state, Eq. \ref{eq:prod}.

\begin{figure}[t]
  \includegraphics[width=6.0cm]{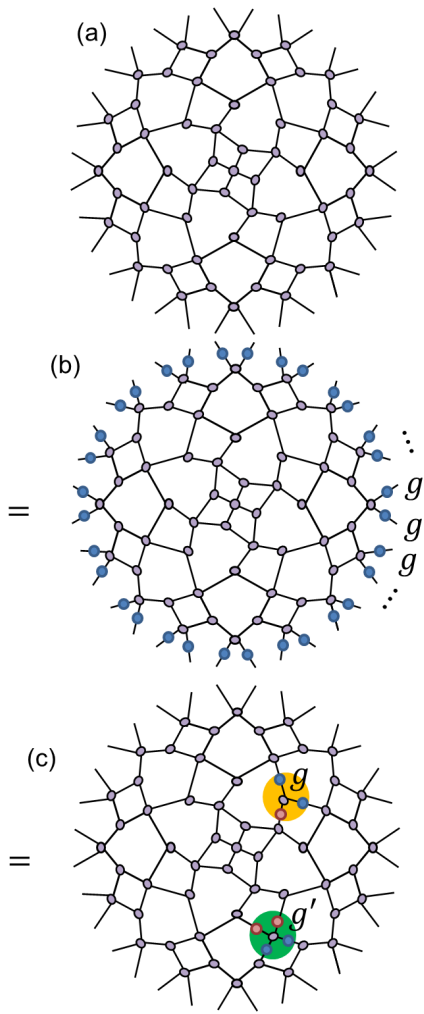}
\caption{
(a) The multi-scale entanglement renormalization ansatz (MERA) for a lattice $\mathcal{L}$ of $L=32$ sites in $D=1$ spatial dimensions. Notice that the MERA expands one dimension more than the lattice $\mathcal{L}$. The radial direction of the MERA corresponds to scale. The MERA can be used to represent the ground state $\ket{\Psi}$ of a quantum critical system, corresponding to a CFT, and the network of tensors is a discrete version of AdS geometry.
(b) If the ground state $\ket{\Psi}$ has a global on-site symmetry $\mathcal{G}$, then the tensor network is invariant under the action of the symmetry simultaneously on all the sites of $\mathcal{L}$.
(c) If the MERA is made of $\mathcal{G}$-symmetric tensors, then not only does it preserves the symmetry globally (lemma 2), but the bulk of the tensor network is also invariant under the action of $\mathcal{G}$ \textit{locally} on individual tensors, where the symmetry acts on all the indices of the tensor at once.
\label{fig:Holo}}
\end{figure}

\section{Discussion}
\label{sec:holo}

In this paper we have considered tensor networks that represent $\mathcal{G}$-symmetric many-body wave-functions. In Sect. \ref{sec:ttn} we have shown that the use of $\mathcal{G}$-symmetric tensors does not require an increase in bond dimension $\chi$ when the tensor network has no loops (that is, a TTN), that is, $\chi_{\nsym}^{\nmin} = \chi^{\nmin}$. Moreover, in Sect. \ref{sec:tnloops} we have described explicit examples of states described with a tensor network with loops (namely with an MPS with PBC, and with the MERA) such that the use of $\mathcal{G}$-symmetric tensors necessarily requires a larger bond dimension $\chi$, so that $\chi_{\nsym}^{\nmin} > \chi^{\nmin}$.

It is important to emphasize that in this second case it is not necessarily true that non-symmetric tensors produce the most compact or computationally convenient tensor network representation. Indeed, the sparse structure of $\mathcal{G}$-symmetric tensors implies that they may depend on less parameters, and result in less costly calculations, even when their bond dimension is larger. Therefore, for tensor networks with loops, establishing whether the use of $\mathcal{G}$-symmetric tensors leads to a more compact and computationally less expensive representation of the many-body wave-function requires a more detailed analysis, and the final conclusion may depend on the specific many-body wave-function that is being represented.

We also note that the discussion of this paper considered only different \textit{exact} representations (with and without $\mathcal{G}$-symmetric tensors) of some fixed $\mathcal{G}$-symmetric many-body wave-function $\ket{\Psi}$. Another context of interest is that of using a tensor network as a variational ansatz. In that case, for a fixed bond dimension $\chi$, we may aim at producing an approximate description of a $\mathcal{G}$-symmetric ground state by minimizing the expectation value of the Hamiltonian. If obtaining the lowest possible variational energy $E_{\chi}$ is the only goal, it is clear that an unconstrained search (where the tensor network is not required to represent a $\mathcal{G}$-symmetric state) is likely to produce a lower energy than a constrained search (where we enforce symmetry constraints to the variational wave-function).
 
Finally, we notice that there are situations where the use of $G$-symmetric tensors appears as mandatory, regardless of computational costs or variational energy considerations. Two of these situations, analyzed in Ref. \cite{SinghVidal}, relate to using the MERA to produce a symmetry-protected renormalization group flow (see also \cite{Chen5}). Here we review them briefly.

First, as we have seen at the end of Sect. \ref{sec:tnloops}, the use of $\mathcal{G}$-symmetric tensors produces a \textit{symmetry-protecting} renormalization group flow that has the $\mathcal{G}$-symmetric state $\ket{\Psi}$ in Eq. \ref{eq:PsiMERA} as a fixed-point; whereas using non-symmetric tensors produces a renormalization group transformation where the same state $\ket{\Psi}$ flows into a product state $\ket{\mbox{prod}}$. It turns out that the state $\ket{\Psi}$ and the product state $\ket{\mbox{prod}}$ belong to different symmetry protected topological (SPT) phases under time-reversal symmetry ($\mathcal{G} = \mathbb{Z}_2^T$) \cite{Pollmann, Chen1, Schuch11, Chen2, Chen3, Chen4}. A symmetry-protecting renormalization group transformation, made of $\mathcal{G}$-symmetric tensors, can be used to find out what SPT phase a given ground state belongs to, by observing what fixed point it flows to. However, if the renormalization group transformation is not made of $\mathcal{G}$-symmetric tensors, then nothing prevents the state of the system from flowing towards the product state $\ket{\mbox{prod}}$, which corresponds to the trivial SPT phase. We conclude \cite{Chen5,SinghVidal} that a $\mathcal{G}$-symmetry protecting coarse-graining transformation, made of $\mathcal{G}$-symmetric tensors, is needed in order to define a renormalization group with the right structure of fixed points -- namely, different fixed-points for different SPT phases.

Second, the use of $\mathcal{G}$-symmetric tensors is also essential in order to reproduce certain properties of the holographic principle with the MERA. Recall that the AdS/CFT correspondence \cite{Maldacena} establishes a connection between a conformal field theory (CFT) and a theory of gravity in anti de Sitter (AdS) space in one additional dimension, which corresponds to scale. In this scenario, there is a dictionary between properties of the CFT (seen as living at the boundary of the space with one additional dimension) and properties of the gravity theory (seen to live in the bulk of AdS space). In particular, a global symmetry of the CFT corresponds to a gauge symmetry of the gravity theory \cite{Witten}. In Refs. \cite{Swingle1,Swingle2, Branching, Metric, Ho1, Ho2, Ho3, Ho4, Ho5}, the MERA is regarded as a lattice realization of the holographic principle. In Ref. \cite{SinghVidal} we pointed out that, if we use $\mathcal{G}$-symmetric tensors to represent a many-body wave-function with a global, on-site symmetry $\mathcal{G}$, then the resulting tensor network, extending in one additional dimension (corresponding to scale) has a gauge symmetry given by group $\mathcal{G}$, see Fig. \ref{fig:Holo}, thus matching a property observed in AdS/CFT \cite{Maldacena, Witten}.

In conclusion, we have investigated the trade-off between using a tensor network with minimal bond dimension and one made of symmetric tensors. We have shown that when describing a $\mathcal{G}$-symmetric many-body wave-function with a TTN, one can always replace non-symmetric tensors with $\mathcal{G}$-symmetric ones without increasing the bond dimension, and that this is no longer the case for tensor networks that contain loops. However, we pointed out that in the latter case there might still be good reasons, both conceptual and computational, to prefer using $\mathcal{G}$-symmetric tensors.

\textbf{Acknowledgements.-} G.V. thanks S. Singh for hospitality, and the Australian Research Council Centre of Excellence for Engineered Quantum Systems. This research was supported in part by Perimeter Institute for Theoretical Physics. Research at Perimeter Institute is supported by the Government of Canada through Industry Canada and by the Province of Ontario through the Ministry of Research and Innovation.


\begin{thebibliography}{74}
\bibitem{nonTrivalentTTN} The proof of Sec.~\ref{ssec:proof} can be readily generalized to the case where tensors have more than three indices as follows. Given a tree tensor network made of tensors with more than three indices, we obtain another TTN that is made of only trivalent tensors by decomposing each higher valence tensor in terms of trivalent tensors. Following the construction employed in the proof, we then replace all tensors in the resulting trivalent TTN with symmetric tensors. Finally, the symmetric tensors that replace the decomposed tensors are contracted back into (symmetric) higher valence tensors.
\bibitem{invariantSVD} By Schur's lemma an invariant matrix $\hat{\rho} : \mathbb{V} \rightarrow \mathbb{V}$ is block diagonal in a representation of the symmetry on the vector space $\mathbb{V}$, $\hat{\rho} = \bigoplus_j \hat{\rho}_j,$ where $j$ labels the irreducible representations of the group. Subsequently, matrix $\hat{\rho}$ can be eigenvalue decomposed $\hat{\rho} = \hat{Q}\hat{D}\hat{Q}^\dagger$, for example, by decomposing each block $\hat{\rho}_j=\hat{Q}_j\hat{D}_j\hat{Q}_j^\dagger$ independently to give output matrices $\hat{Q}$ and $\hat{D}$ that are also block diagonal in the representation,$\hat{Q} = \bigoplus_j \hat{Q}_j$ and $\hat{D} = \bigoplus_j \hat{D}_j$, and are therefore invariant, in accordance with Schur's Lemma.
\bibitem{TimeRev} Each qubit can be thought of as a spin $1/2$ degree of freedom, with a spin operator $\vec{S}$ that behaves analogously to the angular momentum operator $\vec{L}\equiv \vec{X}\times \vec{P}$. Under time reversal $t \rightarrow -t$, which transforms position and momentum operators according to $\vec{X} \rightarrow \vec{X}$, $\vec{P} \rightarrow -\vec{P}$, the angular momentum operator $\vec{L}$ reverses its direction, $\vec{L} \rightarrow -\vec{L}$. Therefore, we would expect that under time reversal the spin operator $\vec{S}$ also reverses its direction, $\vec{S} \rightarrow -\vec{S}$. One can see that, indeed, $\mathcal{T}$ acting on the state of a qubit reverses its direction in the Bloch sphere.
\end{thebibliography}
\end{document}